\documentstyle[titlepage,preprint,aps,floats,epsfig]{revtex}
\newcommand{\beq}{\begin{equation}}
\newcommand{\eeq}{\end{equation}}
\newcommand{\eq}[1]{Eq.(\ref{#1})}

\begin{document}

\draft
\tighten
\title {Radiative-Recoil Corrections of Order $\alpha(Z\alpha)^5(m/M)m$
to Lamb Shift Revisited}
\medskip
\author {Michael I. Eides \thanks{E-mail
address:  eides@pa.uky.edu, eides@thd.pnpi.spb.ru}}
\address{Department of Physics and Astronomy, University of Kentucky,
Lexington, KY 40506, USA \\
and Petersburg Nuclear Physics Institute, Gatchina,
St.Petersburg 188350, Russia}
\author{Howard Grotch\thanks{E-mail address: asdean@pop.uky.edu}}
\address{Department of Physics and Astronomy,
University of Kentucky, Lexington, KY 40506, USA}
\author{Valery A. Shelyuto \thanks{E-mail address: shelyuto@vniim.ru}}
\address{D. I. Mendeleev Institute of
Metrology, St.Petersburg 198005, Russia}
\date{December, 2000}

\maketitle

\begin{abstract}
The results and main steps of an analytic calculation of
radiative-recoil corrections of order $\alpha(Z\alpha)^5(m/M)m$ to the
Lamb shift in hydrogen are presented. The calculations are performed in
the infrared safe Yennie gauge. The discrepancy between two previous
numerical calculations of these corrections existing in the literature
is resolved. Our new result eliminates the largest source of the
theoretical uncertainty in the magnitude of the deuterium-hydrogen
isotope shift.
\end{abstract}

\section{Introduction}

The spectacular experimental progress in precise measurements of the
energy levels in light hydrogenlike atoms achieved in recent years was
matched by an equally impressive theoretical developments (see, e.g.,
the review in \cite{review}, and references therein). Still, there is
a number of unsolved theoretical problems, not least among them the
magnitude of radiative-recoil corrections to the Lamb shift of order
$\alpha(Z\alpha)^5(\lowercase{m}/M)m$. This is the first nontrivial
radiative-recoil correction, and it is generated by the radiative
insertions in the exchanged photon lines and in the electron line. The
correction generated by the one-loop polarization insertions in the
exchanged photon lines was independently calculated analytically in
\cite{egradrecoil} and \cite{pach95}. The results of these works
are in excellent agreement. Radiative-recoil corrections generated by
radiative insertions in the electron line were obtained numerically in
\cite{bg85,bg871,bg87} and in \cite{pach95}. The results of these
works contradict each other. In this work we present an analytic
calculation of the radiative-recoil corrections of order
$\alpha(Z\alpha)^5(\lowercase{m}/M)m$, and resolve the discrepancy
between different theoretical results.

\begin{figure}[ht]
\centerline{\epsfig{file=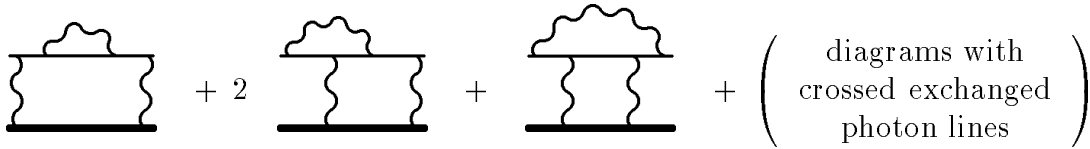,height=2cm}}
\vspace{0.5cm}
\caption{Electron-line radiative-recoil corrections}
\label{ellineradreclamb}
\end{figure}

Calculation of the contributions of order $\alpha(Z\alpha)^5m$ to the
Lamb shift is greatly facilitated by the applicability of the
scattering approximation. These corrections are generated by
the diagrams with at least two photon exchanges in Fig.\
\ref{ellineradreclamb}, and naively one could expect that the diagrams
with larger number of exchanges are also relevant. However, this does
not happen. First, one has to realize that for high exchanged momenta
expansion in $Z\alpha$ is valid, and addition of any extra exchanged
photon always produces an extra power of $Z\alpha$. Hence, in the
high-momentum region only diagrams with two exchanged photons are
relevant. Contribution of the low-momentum region is suppressed because
the infrared behavior of any radiatively corrected Feynman diagram (or
more accurately any gauge invariant sum of Feynman diagrams) is milder
than the behavior of the skeleton diagram. Hence, unlike the leading
contribution to the Lamb shift, diagrams with higher number of photon
exchanges do not contribute to the corrections of order
$\alpha(Z\alpha)^5m$, and it is sufficient to calculate only the
contributions of the diagrams in Fig.\ \ref{ellineradreclamb} in the
scattering approximation (for more detail, see, e.g., \cite{review}).

In the scattering approximation contribution to the energy shift
generated by the diagrams in Fig.\ \ref{ellineradreclamb} is given by
the integral

\beq  \label{general}
\Delta E ~~=~~-~ \frac{(Z\alpha)^5}{\pi n^3}\:{m_r}^3
\:\int {\frac{d^4 k}{i\pi^2 k^4}} \:
\frac{1}{4} Tr \Bigl[(1 + \gamma_0 )L_{\mu \nu} \Bigr]\:
\frac{1}{4} Tr \Bigl[(1 + \gamma_0 )H_{\mu \nu} \Bigr]~\delta_{l0}~~,
\eeq

\noindent
where $m$ and $m_r$ are the electron and reduced masses,
respectively, $k$ is the momentum of the exchanged photon,
$L_{\mu \nu}$ and $H_{\mu \nu}$ are the electron and the
proton factors, respectively, and the Kronecker symbol $\delta_{l0}$
reminds that the radiative-recoil corrections of order
$\alpha(Z\alpha)^5(\lowercase{m}/M)m$  are different from zero only for
the $S$-states.

The electron factor is equal to the sum of the self-energy, vertex, and
spanning photon insertions in the electron line

\beq
L_{\mu \nu} ~=~ L_{\mu \nu}^{\Sigma} ~+~ 2L_{\mu \nu}^{\Lambda}
~+~ L_{\mu \nu}^{\Xi} ~~,
\eeq

\noindent
and the heavy-line factor is given by the expression

\beq
H_{\mu \nu} ~=~\gamma_{\mu} \frac{\hat{P} + \hat{k} + M}
{k^2 + 2Mk_0 + i0} \gamma_{\nu}~~+~~
\gamma_{\nu}  \frac{\hat{P} - \hat{k} + M}{k^2 - 2Mk_0 +
i0}\gamma_{\mu}~~,
\eeq

\noindent
where $P=(M,{\bf 0})$ is the momentum of the proton.

The expression for the energy shift in \eq{general} contains both
recoil and nonrecoil contributions of order
$\alpha(Z\alpha)^5m$. Nonrecoil correction of this
order is well known from the  early days of quantum electrodynamics,
and in order to simplify further calculations we would like to obtain
an expression for the energy shift free of such contributions. Let us
notice to this end that the characteristic integration momenta in
\eq{general} are of order of the electron mass, the lower momenta are
suppressed by the radiative insertions in electron line, and the
momenta of order of the heavy mass are suppressed by the high power of
the integration momenta in the denominators. Suppression of the high
integration momenta means that the radiative-recoil correction of order
$\alpha(Z\alpha)^5(\lowercase{m}/M)m$ does not contain logarithms of
the mass ratio $\ln(m/M)$ which could originate only from the wide
integration region between the electron and proton masses $m\ll k\ll
M$. Then we can kill all nonrecoil contributions and extract the first
order in the mass ratio contribution simply by differentiating the
integral in \eq{general} over the heavy mass $M$, and letting this
mass to go to infinity afterwards. The integral in \eq{general}
contains the heavy mass only in the heavy particle factor, and to
extract the linear in the mass ratio term we make in the integrand the
substitution

\[
\frac{1}{4} Tr \Bigl[(1 + \gamma_0 )H_{\mu \nu} \Bigr] ~~
\longrightarrow
\]
\[
-M \frac{\partial}{\partial M} ~ \Biggl\{~
\frac{1}{4} ~Tr \Biggl\{~ (1 + \gamma_0 ) \Biggl[~ \gamma_{\mu}
\frac{\hat{P} - \hat{k} + M}{k^2 - 2Mk_0 + i0}\gamma_{\nu}
~~+~~\gamma_{\nu} \frac{\hat{P} + \hat{k} + M}
{k^2 + 2Mk_0 + i0} \gamma_{\mu}~\Biggr] ~\Biggr\} ~\Biggr\}
\]
\beq  \label{epslim}
\longrightarrow
~~ -~ \frac{1}{M}  \: \Bigl\{~k^2 \: g_{\mu 0} g_{\nu 0}
~~-~~ k_0 \:\bigl(~g_{\mu 0} k_{\nu} ~+~ g_{\nu 0} k_{\mu} \bigr)
~~+~~ k_0^2 \: g_{\mu \nu} ~\Bigr\}~\:
\frac{k_0^2 + \frac{ k^4}{4M^2}}{(k_0^2 - \frac{ k^4}{4M^2})^2}.
\eeq

\noindent
Due to the explicit factor $1/M$ before the braces, it is sufficient to
consider the last factor in this expression only in limit  $k/M\to0$
(we remind that the characteristic integration momenta are much less
than the proton mass $k\ll M$). Then the coefficients before the tensor
structures in the last line in \eq{epslim} simplify dramatically.

We are going to calculate the Feynman integrals in the
polar coordinates in the four-dimensional euclidean space, where the
vector $k$ is parametrized in the form $k_0=k\cos\theta$ and $|{\bf
k}|=k\sin\theta$. It is easy to see that the integrals with the
coefficient  before the first tensor structure in the braces in
\eq{epslim} should be calculated as principal value integrals over the
polar angle

\beq
k^2 \: \lim\limits_{\frac{k}{M} \to 0} ~
\frac{k_0^2 + \frac{ k^4}{4M^2}}{(k_0^2 - \frac{ k^4}{4M^2})^2} ~~\to ~~
~~ \lim\limits_{\varepsilon \to 0} ~
\frac{\cos^2{\theta} - \varepsilon^2}{(\cos^2{\theta} + \varepsilon^2)^2}
~~ = ~~ \wp \Bigl(\frac{1}{\cos^2{\theta}}\Bigr) ~~,
\eeq

\noindent
where $\varepsilon=k/2M$ and $~\wp~$ is the principal value symbol.

The second term in the braces in \eq{epslim} is odd in $k_0$, and it
will give nonzero contribution to the integral over $k$ only multiplied
by another odd in $k_0$ term from the electron factor.  But this means
that the electron factor would effectively supply an extra power of
$k_0$ in the numerator, and we can safely take the limit
$\varepsilon\to0$ in the coefficient before the second tensor structure
in \eq{epslim}.  The factor before the $g_{mu\nu}$ term in \eq{epslim}
admits a smooth limit for $\varepsilon\to0$. Then the heavy particle
factor turns into

\beq
\frac{1}{4} Tr \Bigl[(1 + \gamma_0 )H_{\mu \nu} \Bigr] ~~\to
~~-~ \frac{1}{M} \: \Bigl\{~k^2 \: g_{\mu 0} g_{\nu 0} \:
\wp \Bigl(\frac{1}{k_0^2}\Bigr)
~~-~~ \bigl(~g_{\mu 0} k_{\nu} ~+~ g_{\nu 0} k_{\mu} \bigr) \:
\frac{1}{k_0}
~~+~~ g_{\mu \nu} ~\Bigr\}~~,
\eeq

\noindent
and the general expression for all radiative-recoil corrections
of order $\alpha(Z\alpha)^5(\lowercase{m}/M)m$  acquires the form

\[
\Delta E ~~=~~\frac{(Z\alpha)^5}{\pi n^3}\:
\frac{m_r^3}{M}
\:\int {\frac{d^4 k}{i\pi^2 k^4}} \:
\frac{1}{4} Tr \biggl\{(1 + \gamma_0 )
\Bigl[ L_{\mu \nu}^{\Sigma} ~+~ 2L_{\mu \nu}^{\Lambda}
~+~ L_{\mu \nu}^{\Xi} \Bigr]\biggr\} ~
\]
\beq \label{ordermm}
~ \Bigl\{~k^2 \: g_{\mu 0} g_{\nu 0}
\: \wp \Bigl(\frac{1}{k_0^2}\Bigr)
~~-~~ \bigl(~g_{\mu 0} k_{\nu} ~+~ g_{\nu 0} k_{\mu} \bigr) \:
\frac{1}{k_0} ~~+~~ g_{\mu \nu} ~\Bigr\}~\delta_{l0}~~.
\eeq

\noindent
This expression is much more convenient for calculations than
\eq{general} because it depends on the heavy mass only
through the explicit factor $m/M$ before the integral.

The expression for the energy shift in \eq{ordermm} is linearly
infrared divergent like $1/\gamma$ where $\gamma$ is an auxiliary
infrared cutoff in integration over $k$. This linear infrared divergence
is the price we have to pay for the simplicity of the scattering
approximation.  The point is that the expressions for the energy shifts
in \eq{general} and \eq{ordermm} contain not only corrections of
relative order $(Z\alpha)^5$ but also the corrections of the previous
order in $Z\alpha$. If we would not ignore small virtualities of the
external electron lines and the external wave functions the naive
infrared divergence would be regularized at the characteristic atomic
scale $\gamma\sim mZ\alpha$. We are not interested in the contributions
of the previous order in $Z\alpha$, and will simply throw  away
linearly infrared divergent contributions in our calculations.
Remaining infrared finite contribution is just the radiative-recoil
correction of relative order $\alpha(Z\alpha)^5$. This strategy works
because the radiative-recoil corrections under consideration do not
contain  logarithms of $Z\alpha$.  However, individual diagrams could
contain logarithms of the infrared cutoff which should cancel in the
final result. This cancellation serves as an additional test of the
correctness of all calculations. In the scattering approximation the
diagrams in Fig.~\ \ref{ellineradreclamb} form a complete gauge
invariant set, and we can use arbitrary gauge for their calculation.
In order to improve the low-momentum behavior of individual diagrams
we use the Yennie gauge for the radiative photons.

\section{Mass Operator Contribution}

Let us consider first the contribution to the radiative-recoil
correction of order $\alpha(Z\alpha)^5(m/M)m$ generated by the diagrams
with the self-energy insertions in the electron line in Fig.\
\ref{ellineradreclamb}. The renormalized mass operator in the Yennie
gauge has the form (see, e.g., \cite{eksann1})

\beq
\Sigma^R (p) ~=~ \frac{\alpha}{4 \pi}\: (\hat{p}-m)^2 \:
\int_0^1 {dx}~\frac{- 3 \, \hat{p}\,x}{m^2 x + (m^2 - p^2)(1-x)}~~.
\eeq

According to \eq{ordermm} respective contribution to the Lamb shift may
be written as

\[
\Delta E_{\,\Sigma} ~~=~~-~ \frac{3}{4}\:\frac{\alpha(Z\alpha)^5}{\pi^2
n^3} \: \frac{m_r^3}{M} \:
\int_0^1 {dx} \: \int {\frac{d^4
k}{i\pi^2}} \: \frac{x}{k^4 \: \Delta_1}  ~\frac{1}{4} ~ Tr
\Bigl[~ (1 + \gamma_0 ) ~\gamma_{\mu} ~ \bigl(~\hat{p} - \hat{k}
~\bigr) ~\gamma_{\nu} ~ \Bigr]~
\]
\[
\Bigl\{~k^2 \: g_{\mu 0} g_{\nu 0} \:
\wp \Bigl(\frac{1}{k_0^2}\Bigr)
~~-~~ \bigl(~g_{\mu 0} k_{\nu} ~+~ g_{\nu 0} k_{\mu} \bigr) \:
\frac{1}{k_0}
~~+~~ g_{\mu \nu} ~\Bigr\}
\]
\beq  \label{massop}
=~-~ \frac{3}{4}\:\frac{\alpha(Z\alpha)^5}{\pi^2 n^3}
\: \frac{m_r^3}{M}
\: \int_0^1 {dx} \: \int {\frac{d^4 k}{i\pi^2}} \:
\frac{x}{k^4 \: \Delta_1} ~
\Bigl[~ -~ 4m + 2k_0
~+~ k^2 \: \frac{1}{k_0}
~+~ m  k^2 \:
\wp \Bigl(\frac{1}{k_0^2}\Bigr) ~\Bigr]~,
\eeq

\noindent
where

\[
\Delta_1 ~=~ m^2 x + 2pk (1-x) - k^2 (1-x) ~\equiv ~
(1-x) \: (- k^2 + 2m k_0 + a^2_1)~~,
\]

\noindent
and $a^2_1={m^2 x}/(1-x)$. In the formulae below we will often use
dimensionless momenta measured in terms of the electron mass $k\to mk$,
and a shorthand notation $C ~=~ \alpha(Z\alpha)^5/(\pi^2
n^3)({m}/{M})({m_r}/{m})^3m$ for the common normalization factor.

As an example of our calculations let us consider evaluation of the
contribution to the energy shift generated by the last
most infrared singular term in the square brackets in \eq{massop}

\[
\Delta E_{\,\Sigma}^{'} ~=~-~  \frac{3C}{4}
\: \int_0^1 {dx} \: \int {\frac{d^4 k}{i\pi^2}} \:
\frac{x}{k^2 \: (-k^2 + 2k_0 + a^2_1)} \:
\wp \Bigl(\frac{1}{k_0^2}\Bigr)
\]
\beq                        \label{infrsingself}
=~-~ \frac{3C}{4} \: \int_0^1 {dx} ~ a^2_1 \:
\int_{\gamma}^{\infty}{\frac{dk^2}{k^2}} \: \frac{2}{\pi}
\int_0^{\pi} {d \theta} ~
\frac{\sin^2{\theta}~(k^2 + a_1^2)}{(k^2 + a_1^2)^2 + 4k^2 \cos^2{\theta}}
\:  \wp \Bigl(\frac{1}{\cos^2{\theta}}\Bigr)~~,
\eeq

\noindent
The integration in the last line goes over the four-dimensional
euclidean space, and we have introduced an auxiliary infrared cutoff
$\gamma$. Using the identity

\beq
\frac{k^2 + a_1^2}{(k^2 + a_1^2)^2 + 4k^2 \cos^2{\theta}}~=~
\frac{1}{k^2 + a_1^2} ~-~
\frac{4k^2 \: \cos^2{\theta}}{(k^2 + a_1^2)\: [~(k^2 + a_1^2)^2
+ 4k^2 \cos^2{\theta}~]}~~,
\eeq

\noindent
we represent the integral in \eq{infrsingself} in the form

\[
\Delta E_{\,\Sigma}^{'}~~=~-~ \frac{3C}{4} \: \int_0^1 {dx} ~ a^2_1
\:  \int_{\gamma}^{\infty}{\frac{dk^2}{k^2}} \: \frac{2}{\pi}
\int_0^{\pi} {d \theta} ~ \sin^2{\theta} \: \biggl\{~
\frac{1}{k^2 + a_1^2} \: \wp \Bigl(\frac{1}{\cos^2{\theta}}\Bigr)
\]
\beq
-~\frac{4k^2}{(k^2 + a_1^2)[~(k^2 + a_1^2)^2
+ 4k^2 \cos^2{\theta}~]}~\biggr\}~~,
\eeq

\noindent
where the principal value integral contains only a trivial
dependence on the angles.

All principal value integrals we need in this work, in particular the
integrals

\[
\frac{2}{\pi} \int_0^{\pi} {d \theta} ~ \:
\sin^2{\theta} \:
\wp \Bigl(\frac{1}{cos^2{\theta}}\Bigr) ~~=~~ -~2 ~~,
\]
\[
\frac{2}{\pi} \int_0^{\pi} {d \theta} ~ \:
\sin^4{\theta} \:
\wp \Bigl(\frac{1}{cos^2{\theta}}\Bigr) ~~=~~ -~3 ~~,
\]

\noindent
may be obtained from the basic integral

\beq
\int_0^{\pi} {d \theta} ~ \:
\wp \Bigl(\frac{1}{cos^2{\theta}}\Bigr) ~~=~~ 0 ~~,
\eeq

\noindent
with the help of algebraic transformations.

Now we can easily complete calculation of the integral in
\eq{infrsingself}

\[
\Delta E_{\,\Sigma}^{'}~= \frac{3C}{4}  \int_0^1 {dx}~
\biggl\{~2 \ln{\frac{a^2_1}{\gamma^2}} ~-~ 2 \ln{\frac{1+a^2_1}{a^2_1}}
~+~ \frac{4}{a} \: \arctan{\frac{1}{a_1}}~\biggr\}
=~ C \biggl\{~3 \ln{\frac{1}{\gamma}} ~+~ \frac{3\pi^2}{8} \biggr\}~~.
\]

Other integrals  in \eq{massop} are calculated in the same way, and we
obtain the total self-energy contribution to the energy shift in the
form

\beq  \label{sigmatot}
\delta E_{\,\Sigma} ~~=~~ \frac{\alpha(Z\alpha)^5}{\pi^2 n^3}
\, \frac{m}{M}\left(\frac{m_r}{m}\right)^3 m\:\biggl[~ 9 \:
\ln{\frac{1}{\gamma}} ~\biggr].
\eeq

\section{Vertex Contribution}

We calculate the contribution to the Lamb shift generated by the vertex
insertion in the electron line in Fig.\ \ref{ellineradreclamb} with the
help of the compact expression for the electron-photon vertex in the
Yennie gauge used in our earlier work on the radiative corrections of
order $\alpha^2(Z\alpha)^5m$ \cite{es}

\beq
\Lambda_{\mu}(p, p-k) ~=~ \frac{\alpha}{4\pi} \:
\sum_{n = 1}^2 \frac{F^{(n)}_{\mu}}{\Delta^n},
\eeq

\noindent
where

\[
F^{(1)}_{\mu} ~~=~~ 3 \gamma_{\mu} \: \Bigr[~ k^2 - 2p \: k
~+~ (2 - x) \: \Delta ~\Bigr]
~~-~~x \: \Bigl[~ 3 \gamma_{\alpha}~(\hat{p} + m)~\gamma_{\mu}
~(\hat{p} - \hat{k} + m)~\gamma^{\alpha}
\]
\[
-~ 6(p - k)\: Q ~\gamma_{\mu}
~+~ \gamma_{\alpha}~ \hat{Q} ~\gamma_{\mu}
~(\hat{p} - \hat{k} + m)~\gamma^{\alpha}
~+~  \gamma_{\alpha}~(\hat{p} + m)~\gamma_{\mu}
~\hat{Q} ~ \gamma^{\alpha}
\]
\beq
+~ 2~\gamma_{\mu}~(\hat{p} - \hat{k} + m)~\hat{Q}
~+~  2 ~\hat{Q} ~(\hat{p} + m)~\gamma_{\mu} ~\Bigr]
~~+~~ x^2 \: \Bigl(~ 2 ~\hat{Q} ~\gamma_{\mu} ~\hat{Q}
~+~ Q^2 ~\gamma_{\mu} ~\Bigr) ~~,
\eeq
\beq
F^{(2)}_{\mu} ~~=~~ 2x(1 - x) ~ \gamma_{\mu}~(\hat{p} - \hat{k} + m)
\: \Bigl[~\hat{Q} ~\hat{p}~ \hat{Q} ~-~ \hat{p}~ Q^2 ~\Bigr]~~,
\eeq
\beq  \label{abdef}
\Delta ~=~ m^2 x + 2pk(1-x)z - k^2z(1-xz)\equiv z(1-xz)(~ a^2 +
2k_0b - k^2~)~,
\eeq

\noindent
and $Q= -p + kz$, $p^2=m^2$.

According to \eq{ordermm} radiative-recoil contribution of
order  $\alpha(Z\alpha)^5(m/M)m$ generated by the vertex insertion has
the form (we use dimensionless integration momenta below)

\[
\Delta E_{\,\Lambda} ~~=~~
2 \: \frac{\alpha(Z\alpha)^5}{\pi^2 n^3} \, \frac{m}{M}
\left(\frac{m_r}{m}\right)^3 m\:
\: \int_0^1 {dx} \: \int_0^1 {dz}
\: \int {\frac{d^4 k}{i\pi^2}} \: \frac{1}{k^4}
\: \frac{1}{k^2 - 2k_0} ~\:
\]
\beq \label{genvert}
\: ~ \sum_{n = 1}^2 \frac{1}{\Delta^n} \:
\Bigl[ ~V_0^{(n)} ~~+ ~~ V_1^{(n)} \: \frac{1}{k_0}
~~+~~ V_2^{(n)} \: \wp \Bigl(\frac{1}{k_0^2}\Bigr)~\Bigr] ~~,
\eeq

\noindent
where

\[
V_0^{(n)} ~~\equiv ~~
\frac{1}{4} ~ Tr \Bigl[~ (1 + \gamma_0 ) ~
F^{(n)}_{\mu}~(\hat{p} - \hat{k} + 1)~\gamma_{\nu} ~\Bigr] \:
g_{\mu \nu} ~~,
\]
\[
V_1^{(n)} ~~\equiv ~~
\frac{1}{4} ~ Tr \Bigl[~ (1 + \gamma_0 ) ~
F^{(n)}_{\mu}~(\hat{p} - \hat{k} + 1)~\gamma_{\nu} ~\Bigr] \:
(~-~g_{\mu 0} k_{\nu} ~-~ g_{\nu 0} k_{\mu} ~) ~~,
\]
\[
V_2^{(n)} ~~\equiv ~~
\frac{1}{4} ~ Tr \Bigl[~ (1 + \gamma_0 ) ~
F^{(n)}_{\mu}~(\hat{p} - \hat{k} + 1)~\gamma_{\nu} ~\Bigr] \:
k^2 \: g_{\mu 0} g_{\nu 0} ~~.
\]

\noindent
Calculating traces and contracting the Lorentz indices
we obtain the numerator factors in the square brackets in \eq{genvert}

\[
V_0^{(1)} ~~+ ~~ V_1^{(1)} \,\frac{1}{k_0}
~~+~~ V_2^{(1)} \, \wp \Bigl(\frac{1}{k_0^2}\Bigr)
\]
\[
= ~~4x(1-x)k^2 \Bigl[~1 ~+~ \frac{1}{k_0} ~\Bigr]
+ ~~2x(5-9z+6z^2+4xz-4xz^2)k^2
\Bigl[~k_0 ~+~ \frac{2}{k_0} ~\Bigr]
\]
\[
+~~(k^2-2k_0) \Bigl\{~\Bigl[~-~6(1-2z) ~+~
2x(-1-10z+2x+2xz) ~\Bigr]
\]
\beq  \label{n1vertex}
+ ~~\Bigl[~6(1-2z) ~+~ 2x(1+z+6z^2+2xz-4xz^2) ~\Bigr]
k^2 \wp \Bigl(\frac{1}{k_0^2}\Bigr)
\eeq
\[
+ ~~\Bigl[~6(1-2z) ~+~ 2x(-5+10z-4xz) ~\Bigr]k_0
+ ~~\Bigl[~3(1 - 2z) ~+~ x(z+6z^2-4xz^2) ~\Bigr]
\frac{k^2}{k_0}~\Bigr\}~~,
\]
\[
V_0^{(2)} ~~+ ~~ V_1^{(2)} \: \frac{1}{k_0}
~~+~~ V_2^{(2)} \: \wp \Bigl(\frac{1}{k_0^2}\Bigr)
\]
\[
=~~ -~4x(1 - x)z \: {\bf k}^2 \: \Bigl\{~ 4(1 - z) \:
\Bigl[~1 ~+~ k^2 \: \wp \Bigl(\frac{1}{k_0^2}\Bigr)~\Bigr]
+~~ 2(1 - z) \: (2k_0^2 - k^2) \: \frac{1}{k_0}
\]
\beq
+~~ (k^2 - 2k_0) \: \Bigl[~-~2z ~+~ 2(1-2z)
\: \frac{1}{k_0} ~~+~~
z k^2 \wp \Bigl(\frac{1}{k_0^2}\Bigr) ~\Bigr]~\Bigr\}~~.
\eeq

As a simple example let us consider calculation of the contribution to
the energy shift generated by the terms in the last line in \eq{n1vertex}

\[
\Delta E_{\,\Lambda}^{'} ~=~ 2C \,\int_0^1 {dx} \int_0^1 {dz}
\int {\frac{d^4 k}{i\pi^2}} \, \frac{1}{k^4 \Delta}
\Biggl\{\Bigl[6(1-2z)+2x(-5+10z-4xz)\Bigr] k_0
\]
\[
+ ~\Bigl[3(1-2z)+x(z+6z^2-4xz^2)\Bigr]
\frac{k^2}{k_0}\Biggr\}
\]
\[
=~ 2C \, \int_0^1 {dx} \int_0^1 {dz} ~\frac{b}{z(1-xz)}
\int_{0}^{\infty} {d k^2}
~\frac{2}{\pi} \int_0^{\pi} {d \theta} ~
\frac{\sin^2{\theta}}{(k^2+a^2)^2 + 4b^2 k^2 \cos^2{\theta}} ~ \:
\]
\beq  \label{vetex1}
\: ~\Biggl\{\Bigl[6(1-2z)
+2x(-5+10z-4xz) \Bigr]\, \cos^2{\theta}
~+~\Bigl[3(1-2z)+x(z+6z^2-4xz^2)\Bigr] \Biggr\}~~.
\eeq

\noindent
The last integral as all other integrals in this paper may be written
as a linear combination of the integrals of the form

\beq \label{standint}
\int_{0}^{\infty} {d k^2}
~\frac{2}{\pi} \int_0^{\pi} {d \theta} ~
\frac{\sin^2{\theta} ~ \cos^{2l}{\theta} ~ (k^2)^m ~ (k^2+a^2)^n}
{[(k^2+a^2)^2 + 4b^2 k^2 \cos^2{\theta}]^p}~~,
\eeq

\noindent
where $l,m,n = -1,0,1$ and $p = 1,2,3$. For $l=-1$ the integrals with
$\cos^2{\theta}$ in the denominator should be interpreted as principal
value integrals. All these integrals may be calculated in terms of four
standard functions of the parameters $a$, $b$

\beq
L_0 = \ln{\frac{a^2+b^2}{a^2}}, ~~~ L_1 = 1 - \frac{a^2}{b^2} L_0,
~~~ L_2 = 1 - \frac{2a^2}{b^2} L_1, ~~~ \frac{b}{a} \arctan{\frac{b}{a}}.
\eeq

The result of integration for our example in \eq{vetex1} is

\[
\Delta E_{\,\Lambda}^{'}~=~
2C \int_0^1 {dx} \int_0^1 {dz} ~\frac{1}{z(1-x)}
~\Biggl\{\Bigl[6(1-2z)+2x(-5+10z-4xz)\Bigr]~
\Bigl(\frac12 L_0 ~-~ \frac12 L_1 \Bigr)
\]
\[
+ ~\Bigl[3(1-2z)+x(z+6z^2-4xz^2)\Bigr] ~
\Bigl(-L_0 ~+~ \frac{2b}{a} \arctan{\frac{b}{a}}\Bigr) \Biggr\}
\]
\beq
=
2 \,\frac{\alpha(Z\alpha)^5}{\pi^2 n^3} \, \frac{m}{M}
\left(\frac{m_r}{m}\right)^3 m\:
\biggl[~-~\frac{11}{2}\: \zeta{(3)}~+~7\pi^2\: \ln{2}
~-~ \frac{53\pi^2}{12} ~+~7 ~~\biggr]~~.
\eeq

In the contribution to the energy shift in \eq{vetex1} the electron
denominator canceled with a similar term in the numerator, and as a
result all denominators were combined with the help of only two Feynman
parameters $x, z$. Let us turn now to a simple example  when such
cancelation does not take place, and we need to introduce a third
Feynman parameter $t$. The contribution generated by the
first term in \eq{n1vertex} may be written in the form

\[
\delta E_{\,\Lambda}^{''}~=~ 8C \, \int_0^1 {dx} \int_0^1 {dz}
\int {\frac{d^4 k}{i\pi^2}} ~
\frac{1}{k^2 (k^2-2k_0)} ~ \frac{x(1-x)}{\Delta}
\]
\beq  \label{thirdpar}
=-8C \int_0^1 {dx} \int_0^1 {dz}(1-x) a^2 \int_0^1 {dt}
\frac{\partial}{\partial a_t^2}
\int_0^{\infty} {dk^2}
\frac{2}{\pi} \int_0^{\pi} {d \theta}  \sin^2{\theta}
\frac{k^2+a_t^2}{(k^2 + a_t^2)^2 + 4b_t^2 k^2 \cos^2{\theta}},
\eeq

\noindent
where $a_t^2 = a^2 t$ and $b_t = 1 - z(1-z)a^2 t$.

The integral over angles and momenta in \eq{thirdpar} is just of the
standard form \eq{standint}, and may be easily calculated. Integration
over $t$ is facilitated by the simple observation that the $t$-integral
may be written as an integral over $a_t^2$ with the upper limit $a^2$.
After this transformation all dependence of the new integral on $x$ is
hidden in this upper integration limit, and we can get rid of the third
Feynman parameter integrating by parts over $x$ (for more details see
\cite{eksann1})

\[
\Delta E_{\,\Lambda}^{''}~=~ 8C \, \int_0^1 {dx}
\int_0^1 {dz}(1-x) a^2
\int_0^1 {\frac{dt}{b_t^2}} ~ L_{0t}
\]
\beq
=~~ 8C \, \int_0^1 {dx} \int_0^1 {dz} (1-x)
\int_0^{a^2} {\frac{da_t^2}{b_t^2}} ~ L_{0t}
=~~ 4C \,\int_0^1 {dx} \int_0^1 {\frac{dz}{z}}~L_{0}
~=~ 2C \,\biggl[\frac{\pi^2}{2} - 2 \biggr].
\eeq

Contributions of the other terms in \eq{n1vertex} are calculated in the
same fashion, and the total contribution of all terms with $n=1$ in
\eq{genvert} is equal to

\beq  \label{vertn1}
\Delta E_{\,\Lambda}^{(1)} ~~=~~
2 \,\frac{\alpha(Z\alpha)^5}{\pi^2 n^3} \, \frac{m}{M}
\left(\frac{m_r}{m}\right)^3 m\:
\biggl[-~9\ln{\frac{1}{\gamma}}
~+~ \frac{3}{8} \,\zeta{(3)} ~+~ \frac{9\pi^2}{4} \ln{2}
~-~ \frac{3\pi^2}{4} ~+~ 4 ~\biggr]~~.
\eeq

For the total  contribution of all terms with $n=2$ in \eq{genvert} we
obtain

\beq     \label{vertn2}
\Delta E_{\,\Lambda}^{(2)} ~~=~~
2 \,\frac{\alpha(Z\alpha)^5}{\pi^2 n^3} \, \frac{m}{M}
\left(\frac{m_r}{m}\right)^3 m\:
\biggl[~\frac{21}{8} \,\zeta{(3)} ~-~ \frac{13\pi^2}{4} \ln{2}
~+~ \frac{9\pi^2}{8} ~-~ \frac{35}{4} ~\biggr]~~.
\eeq

Total vertex insertion contribution to the radiative-recoil correction
of order $\alpha(Z\alpha)^5(m/M)m$ is given by the sum of
the contributions in \eq{vertn1} and \eq{vertn2}

\beq  \label{vertextot}
\Delta E_{\,\Lambda} ~~=~~
\frac{\alpha(Z\alpha)^5}{\pi^2 n^3} \, \frac{m}{M}
\left(\frac{m_r}{m}\right)^3 m\:
\biggl[~-~18 \ln{\frac{1}{\gamma}}
~+~ 6 \, \zeta{(3)} ~-~ 2\pi^2 \ln{2} ~+~ \frac{3\pi^2}{4}
~-~ \frac{19}{2} ~\biggr].
\eeq

\section{Spanning Photon Contribution}

The contribution to the Lamb shift generated by the spanning photon
insertion in the electron line in Fig.\ \ref{ellineradreclamb} is
calculated with the help of an explicit expression for the jellyfish
diagram. The small $k$ behavior of the jellyfish diagram
is one of our primary concerns in further calculations, since the
contribution of the previous order is connected just with this
infrared region. The jellyfish diagram is finite at small $k$ in the
Yennie gauge, and this is one of the reasons why we are working in this
gauge. We need such representation for the jellyfish diagrams where
not only the diagram as a whole, but all entries are finite at
$k=0$. The compact expression for the jellyfish diagram with such
properties was used in our earlier work on the radiative corrections of
order $\alpha^2(Z\alpha)^5m$ \cite{es}

\beq
L^{\Xi}_{\mu\nu}=\frac{\alpha}{4\pi}\int_0^1dx\int_0^1dzx(1-z)\sum_{n=1}^3
\frac{M_{\mu\nu}^{(n)}}{\Delta^n},
\eeq

\noindent
where

\beq             \label{xifactor}
M_{\mu\nu}^{(1)}=-2N_{\mu\nu}^{(a)},
\eeq
\[
M_{\mu\nu}^{(2)}=N_{\mu\nu}^{(b)}+2(1-x)N_{\mu\nu}^{(c)}+3N_{\mu\nu}^{(sing)},
\]
\[
M_{\mu\nu}^{(3)}=-2(1-x)N_{\mu\nu}^{(d)},
\]
\[
N_{\mu\nu}^{(a)}=\gamma_\mu(5\hat p-3\hat k)\gamma_\nu+4mg_{\mu\nu}
+x[\hat Q\gamma_\nu\gamma_\mu+\gamma_\nu\gamma_\mu\hat Q+4\gamma_\mu\hat
Q\gamma_\nu],
\]
\beq                      \label{xin}
{N}_{\mu\nu}^{(b)}=2m[\hat Q\gamma_\mu(\hat p+x\hat Q-\hat k+m)
\gamma_\nu+\gamma_\mu(\hat p+x\hat Q-\hat k+m)\gamma_\nu\hat Q]
\eeq
\[
-12(1-2x)m^2\gamma_\mu\hat Q\gamma_\nu
-2x\hat Q\gamma_\nu(\hat p+x\hat Q-\hat k)
\gamma_\mu\hat Q+4xmQ^2g_{\mu\nu}
\]
\[
+[2xQ^2+8(pQ)]\gamma_\mu(\hat p+x\hat Q-\hat k+m)\gamma_\nu,
\]
\[
N_{\mu\nu}^{(c)}=4(pQ)\gamma_\mu\hat p\gamma_\nu+2m^2\gamma_\mu\hat
Q\gamma_\nu,
\]
\[
{N}_{\mu\nu}^{(d)}=8[(pQ)^2-m^2Q^2]\gamma_\mu(\hat p+x\hat Q-\hat
k+m)\gamma_\nu,
\]
\[
N_{\mu\nu}^{(sing)}=4m^2\gamma_\mu(\hat p-\hat k+m)\gamma_\nu ~~,
\]

\noindent
and $a^2$, $b$ and $Q$ where defined in \eq{abdef}. Notice that all
$M_{\mu\nu}^{(i)}$ are infrared finite even at $k=0$.

General expression for the energy shift induced by the spanning photon
insertion has the form  (see \eq{ordermm})

\[
\Delta E_{\,\Xi} ~~=~~
\frac{\alpha(Z\alpha)^5}{\pi^2 n^3} \, \frac{m}{M}
\left(\frac{m_r}{m}\right)^3 m\:
\, \int_0^1 {dx} \int_0^1 {dz}  x(1-z)
\int {\frac{d^4 k}{i\pi^2}} \, \frac{1}{k^4} ~\:
\]
\beq  \label{spann}
\: ~ \sum_{n = 1}^3 \frac{1}{\Delta^n} \,
\Bigl[ ~T_0^{(n)} ~~+ ~~ T_1^{(n)} \, \frac{1}{k_0}
~~+~~ T_2^{(n)} \, \wp \Bigl(\frac{1}{k_0^2}\Bigr)~\Bigr] ~~,
\eeq

\noindent
where

\[
T_0^{(n)} ~~\equiv ~~
\frac{1}{4} ~ Tr \Bigl[~ (1 + \gamma_0 ) ~
M^{(n)}_{\mu \nu}~\Bigr] \:
g_{\mu \nu} ~~,
\]

\[
T_1^{(n)} ~~\equiv ~~
\frac{1}{4} ~ Tr \Bigl[~ (1 + \gamma_0 ) ~
M^{(n)}_{\mu \nu}~\Bigr] \:
(~-~g_{\mu 0} k_{\nu} ~-~ g_{\nu 0} k_{\mu} ~) ~~,
\]

\beq
T_2^{(n)} ~~\equiv ~~
\frac{1}{4} ~ Tr \Bigl[~ (1 + \gamma_0 ) ~
M^{(n)}_{\mu \nu}~\Bigr] \:
k^2 \: g_{\mu 0} g_{\nu 0} ~~.
\eeq

\noindent
Calculating traces and contracting the Lorentz indices
we obtain the numerator factors in the square brackets in \eq{spann}

\[
T_0^{(1)} ~~+ ~~ T_1^{(1)} \: \frac{1}{k_0}
~~+~~ T_2^{(1)} \: \wp \Bigl(\frac{1}{k_0^2}\Bigr)
\]
\[
=~~ 24\:(1 - x) \:  ~+~ 4 \: ( - 3 + 2xz) \: k_0  ~~+~~
2\:(- 3 + 2xz) \:  \frac{k^2}{k_0}
\]
\beq
~~+~~ 6 \: ( - 3 + 2x) \:  k^2 \:
\wp \Bigl(\frac{1}{k_0^2}\Bigr) ~~,
\eeq
\[
T_0^{(2)} ~~+ ~~ T_1^{(2)} \: \frac{1}{k_0}
~~+~~ T_2^{(2)} \: \wp \Bigl(\frac{1}{k_0^2}\Bigr)
\]
\[
=~~ 12x \:   ~~-~~ 8z(1 - x)(4 - x) \:  k_0
~~+~~ 8 z (2 - x - xz) \:  k_0^2
\]
\[
+~~ 4 z (4 + x - 4xz + 2x^2 z) \:  k^2
~~+~~ 8z(1 - x)(4 - x)  \: \frac{k^2}{k_0}
\]
\beq
+~~ 6x \:  k^2 \: \wp \Bigl(\frac{1}{k_0^2}\Bigr)
~~+~~2 z (2 + 2x + 3xz - 4x^2 z) \:  k^4 \:
\wp \Bigl(\frac{1}{k_0^2}\Bigr) ~~,
\eeq

\[
T_0^{(3)} ~~+ ~~ T_1^{(3)} \: \frac{1}{k_0}
~~+~~ T_2^{(3)} \: \wp \Bigl(\frac{1}{k_0^2}\Bigr)
\]

\[
=~~-~16 (1 - x) z^2  {\bf k}^2 \: \Bigl[~
2(- 1 + 2x)  ~+~ 2 (1 - xz) \: k_0
\]

\beq
~~+~~ (1 - xz) \:   \frac{k^2}{k_0}
~~+~~ (2 - x) \:  k^2 \:
\wp \Bigl(\frac{1}{k_0^2}\Bigr) ~\Bigr]~~.
\eeq

We will use the integral

\beq
\Delta E_{\,\Xi}^{'} ~=~
16 C \,\int_0^1 {dx} \int_0^1 {dz} z(1-z)
\int {\frac{d^4 k}{i\pi^2}} ~ \frac{{\bf k}^2}{k^2}~
\frac{-2xz}{\Delta^3} ~
\wp \Bigl(\frac{1}{k_0^2}\Bigr),
\eeq

\noindent
describing one of the contributions with $n=3$ in order to illustrate
one more subtlety encountered in our calculations.  This infrared
divergent integral, contains in the integrand the term $x/\Delta^3$. At
small $k \to 0$ the denominator $\Delta \to x$, and integration over
$x$ becomes too singular. The singular factor $\wp (1/k_0^2)$ in the
integrand  makes things even worse, and we risk to end up with a
divergent integral over the Feynman parameter $x$, instead of an
infrared divergent integral over the momentum.  While linearly infrared
divergent integrals over $k$ have a transparent physical interpretation
as contributions of the previous order in $Z\alpha$, divergent
integrals over $x$ could lead to uncontrollable contributions. We
separate the infrared divergent part of the momentum integral with the
help of the identity

\beq
-2z\int_0^1dx\frac{x}{\Delta^3}=
\frac{1}{k^2-2k_0}+\frac{2z}{1-k^2z(1-z)}+\frac{z^2(1-z)k^2}
{[1-k^2z(1-z)]^2}
\eeq
\[
-z(1-Q^2)\int_0^1dx\left(\frac{1}{\Delta^2}+\frac{2x}{\Delta^3}\right).
\]

\noindent
Then the singular integration over momentum decouples, and we easily
obtain

\beq
\Delta E_{\,\Xi}^{'} ~=~ 16 C \,
\Biggl[\frac{4}{3\gamma} - \frac34~\Biggl].
\eeq

After tedious calculations we obtain

\beq
\Delta E_{\,\Xi}^{(1)} ~~=~~
\frac{\alpha(Z\alpha)^5}{\pi^2 n^3} \, \frac{m}{M}
\left(\frac{m_r}{m}\right)^3 m\:
\biggl[~9\ln{\frac{1}{\gamma}}
~-~ \frac{123}{4} \,\zeta{(3)} ~+~ \frac{3\pi^2}{2} \ln{2}
~-~ \frac{15\pi^2}{16} ~+~ \frac{159}{4} ~\biggr]~~,
\eeq
\beq
\Delta E_{\,\Xi}^{(2)} ~~=~~
\frac{\alpha(Z\alpha)^5}{\pi^2 n^3} \, \frac{m}{M}
\left(\frac{m_r}{m}\right)^3 m\:
\biggl[~12 \,\zeta{(3)}
~+~ \frac{3\pi^2}{8} ~-~ 16 ~\biggr]~~,
\eeq
\beq
\Delta E_{\,\Xi}^{(3)} ~~=~~
\frac{\alpha(Z\alpha)^5}{\pi^2 n^3} \, \frac{m}{M}
\left(\frac{m_r}{m}\right)^3 m\:
\biggl[~\frac{75}{4} \,\zeta{(3)} ~-~ \frac{3\pi^2}{2} \ln{2}
~+~ \frac{9\pi^2}{16} ~-~ \frac{113}{4} ~\biggr]~~,
\eeq

\noindent
for the contributions with $n=1,2,3$ in \eq{spann}, respectively. Let
us emphasize once again that we have thrown away linearly infrared
divergent term $1/\gamma$ in $\Delta E_{\,\Xi}^{(3)}$, which
corresponds to the contribution of the previos order, but preserved all
logarithmically divergent terms which should cancel automatically in
the final result for the energy shift.

Total contribution to the energy shift generated by the spanning photon
insertion in Fig.~\ \ref{ellineradreclamb} is equal to

\beq    \label{jellyfish}
\delta E_{\,\Xi} ~~=~~
\frac{\alpha(Z\alpha)^5}{\pi^2 n^3} \, \frac{m}{M}
\left(\frac{m_r}{m}\right)^3 m\:
\biggl[~9 \ln{\frac{1}{\gamma}} ~-~ \frac{9}{2} ~\biggr].
\eeq

\section{~~~Summary}

Collecting all contributions to the energy shift in \eq{sigmatot},
\eq{vertextot}, and \eq{jellyfish} we see that all logarithmically
infrared divergent contributions cancel in the sum, and obtain the total
radiative-recoil correction of order $\alpha(Z\alpha)^5(m/M)m$

\beq   \label{totalresult}
\Delta E ~~=~~\left(6\zeta(3)-2\pi^2\ln2+\frac{3\pi^2}{4}-14\right)
\frac{\alpha(Z\alpha)^5}{\pi^2 n^3}\:  \frac{m}{M}
\left(\frac{m_r}{m}\right)^3m
\eeq
\[
\approx-13.0676322\ldots
\frac{\alpha(Z\alpha)^5}{\pi^2 n^3}\:  \frac{m}{M}
\left(\frac{m_r}{m}\right)^3m.
\]

\noindent
This result is in excellent agreement with the numerical result in
\cite{pach95}, and this resolves the long standing discrepancy on the
magnitude of the radiative-recoil corrections of order
$\alpha(Z\alpha)^5(m/M)m$ to the Lamb shift. When this paper was in
preparation we learned that the same result was just obtained in
the NRQED framework by Czarnecki and Melnikov \cite{cm}.

Numerically, the correction in \eq{totalresult} contributes

\beq
\Delta E (1S) ~=~~-13.43~\mbox{kHz}
\eeq

\noindent
to the $1S$ Lamb shift in the ground state of hydrogen.
The discrepancy between the theoretical predictions for the $1S$ Lamb
shift calculated  according to \cite{bg85,bg871,bg87} and \cite{pach95}
is about 6 kHz. This discrepancy is not too important for the $1S$ Lamb
shift measurements, since the error bars of even the best current
experimental results are still a few times larger (see, e.g., review in
\cite{review}). What is much more important from the phenomenological
point of view, the radiative-recoil correction is linear in the
electron-nucleus mass ratio, and it directly contributes to the
hydrogen-deuterium isotope shift. The discrepancy between the
theoretical values of the isotope shift calculated  according to
\cite{bg85,bg871,bg87} and \cite{pach95} is about $18$ times
larger than the experimental uncertainty $0.15$ kHz of the isotope
shift \cite{huber}. Thus the analytic result in \eq{totalresult}
eliminates the largest source of the theoretical uncertainty in the
magnitude of the deuterium-hydrogen isotope shift. One can use the
new radiative-recoil correction, and the latest experimental data in
order to obtain a new value for the difference of charge radii
squared of the deuteron and proton, but we will not enter in the
detailed discussion of the phenomenological implications here, since
they were exhaustively discussed in our recent review \cite{review}.

\acknowledgements

We are grateful to A. Czarnecki and K. Melnikov for communicating us
their results prior to publication.

This work was supported by the NSF grant PHY-0049059.


\begin{thebibliography}{99}


\bibitem{review} M. I. Eides, H. Grotch, and V. A. Shelyuto, preprint
PSU/TH/226, hep-ph/0002158, February 2000, Physics Reports, in print.

\bibitem{egradrecoil} M. I. Eides and H. Grotch, Phys. Rev. {\bf A52} (1995)
1757.

\bibitem{pach95} K. Pachucki, Phys. Rev. A {\bf 52}, 1079 (1995).


\bibitem{bg85} G. Bhatt and  H. Grotch, Phys. Rev. A {\bf 31}, 2794 (1985).

\bibitem{bg871} G. Bhatt and  H. Grotch, Phys. Rev. Lett. {\bf 58},
471(1987).

\bibitem{bg87} G. Bhatt and  H. Grotch, Ann. Phys. (NY) {\bf 178}, 1 (1987).

\bibitem{eksann1} M. I. Eides, S. G. Karshenboim, and V. A. Shelyuto, Ann.
Phys.  (NY) {\bf 205} (1991) 231.

\bibitem{es} M. I. Eides and V. A. Shelyuto, Pis'ma Zh. Eksp. Teor. Fiz.
{\bf 61} (1995) 465 [JETP Letters {\bf 61} (1995) 478]; Phys. Rev. {\bf
A52} (1995) 954.

\bibitem{cm} A. Czarnecki and K. Melnikov, preprint SLAC-PUB-8730,
hep-ph/0012053, December 2000.

\bibitem{huber} A. Huber, Th. Udem, B. Gross et al.,
Phys.  Rev. Lett. {\bf 80} 468 (1998).


\end{thebibliography}
\end{document}